\documentclass[pre,twocolumn,showpacs,showkeys]{revtex4}
\usepackage{amsmath,amsgen,amstext,amsbsy,amsopn,amsthm,amssymb,epsf}
\usepackage{graphics,graphicx,wrapfig,epsf}

\usepackage{tikz-cd}
\usetikzlibrary{matrix,arrows,decorations.pathmorphing,positioning}

\def\be{\begin{equation}}
\def\ee{\end{equation}}
\def\ba{\begin{align}}
\def\ea{\end{align}}
\def\bm{\begin{multline}}
\def\eem{\end{mutline}}
\def\bfig{\begin{figure}[htb]}
\def\efig{\end{figure}}

\renewcommand{\leq}{\;\leqslant\;}
\renewcommand{\geq}{\;\geqslant\;}
\newcommand{\dd}{{\rm d}}
\newcommand{\e}[1]{\,{\rm e}^{#1}\,}
\newcommand{\ii}{{\rm i}}

\newcommand{\equal}[1]{\; \substack{#1 \\ {\displaystyle =} \\ \phantom{#1}} \;}

\def\Tr{{\operatorname{Tr\,}}}

\newcommand{\upchi}{\raise 2pt \hbox{$\chi$}}

\def\tra{\lower 2mm \hbox{\includegraphics[width=4mm]{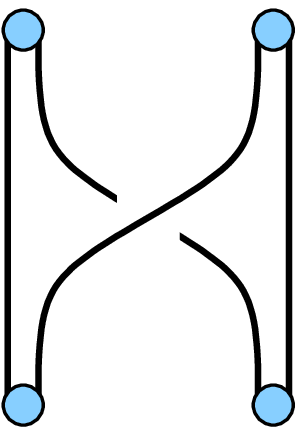}}}
\def\trb{\lower 2mm \hbox{\includegraphics[width=4mm]{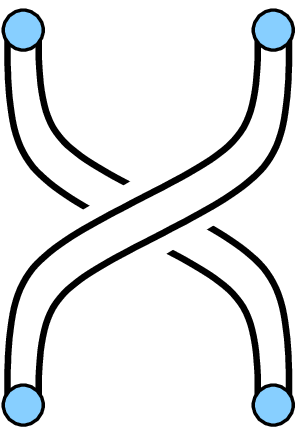}}}
\def\trc{\lower 2mm \hbox{\includegraphics[width=4mm]{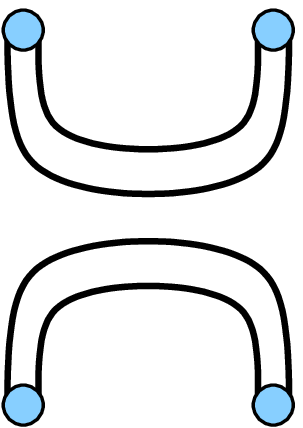}}}
\def\trd{\lower 2mm \hbox{\includegraphics[width=4mm]{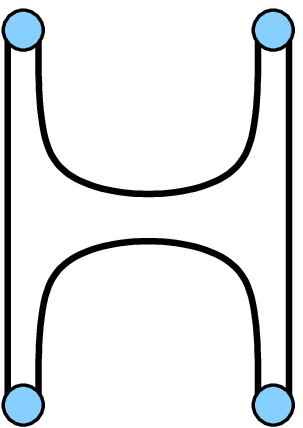}}}
\def\tre{\lower 2mm \hbox{\includegraphics[width=4mm]{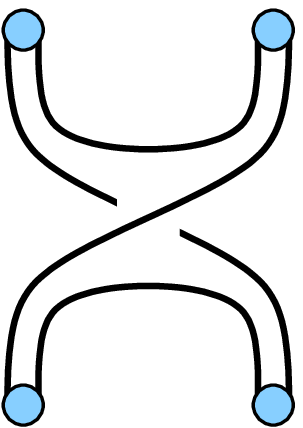}}}

\newcommand{\caH}{{\mathcal H}}\newcommand{\caL}{{\mathcal L}}\newcommand{\caS}{{\mathcal S}}

\def\bbone{{\mathchoice {\rm 1\mskip-4mu l} {\rm 1\mskip-4mu l} {\rm 1\mskip-4.5mu l} {\rm 1\mskip-5mu l}}}

\newcommand{\bbC}{{\mathbb C}}\newcommand{\bbE}{{\mathbb E}}\newcommand{\bbP}{{\mathbb P}}\newcommand{\bbR}{{\mathbb R}}\newcommand{\bbS}{{\mathbb S}}\newcommand{\bbZ}{{\mathbb Z}}

\begin{document}


\title{Ferromagnetism, antiferromagnetism, and the curious nematic phase of $S=1$ quantum spin systems}

\author{Daniel Ueltschi}
\affiliation{Department of Mathematics, University of Warwick, Coventry, CV4 7AL, United Kingdom}
\email{daniel@ueltschi.org}

\begin{abstract}
We investigate the phase diagram of $S=1$ quantum spin systems with SU(2)-invariant interactions, at low temperatures and in three spatial dimensions. Symmetry breaking and the nature of extremal states can be studied using random loop representations. The latter confirm the occurrence of ferro- and antiferromagnetic transitions and the breaking of SU(3) invariance. And they reveal the peculiar nature of the nematic extremal states which {\it minimize} $\sum_{x} (S_{x}^{i})^{2}$.
\end{abstract}


\maketitle


\section{Introduction}
\label{sec intro}

A fascinating aspect of phase transitions is the notion of symmetry breaking. The set of equilibrium states (infinite-volume Gibbs states) of the system possesses the same symmetries as the Hamiltonian. There is typically a unique Gibbs state at high temperatures which is symmetric. But there may be many different Gibbs states at low temperatures, that are not symmetric. An arbitrary state has a unique decomposition in {\it extremal states}, where extremal states are characterised by their clustering properties (decay of truncated correlation functions) and by the fact that they cannot be decomposed in other states.

The decomposition in extremal states is well described in models such as Ising, where the spin-flip symmetry is broken at low temperatures in dimensions two and higher. Periodic equilibrium states can be written as the convex combination of exactly two periodic extremal states. A similar behavior is expected in models with continuous symmetry, such as the Heisenberg and the XY models. In the case of the classical XY model, there are rigorous results due to Pfister and Fr\"ohlich \cite{Pfi1,FP,Pfi2}.

In this article, we study spin 1 quantum systems with SU(2)-invariant interactions. Let $\Lambda \subset \bbZ^{d}$ be a finite lattice, and let $\caH_{\Lambda} = \otimes_{x\in\Lambda} \bbC^{3}$ be the Hilbert space of the system. The Hamiltonian is
\be
H_{\Lambda} = -\sum_{\langle x,y \rangle} \bigl( J_{1} \vec S_{x} \cdot \vec S_{y} + J_{2} (\vec S_{x} \cdot \vec S_{y})^{2} \bigr),
\ee
where the sum is over nearest-neighbors $x,y \in \Lambda$; the spin operators $S_{x}^{i}$, $x\in\Lambda$, $i=1,2,3$ satisfy $[S_{x}^{1}, S_{y}^{2}] = \ii \delta_{x,y} S_{x}^{3}$ (and further cyclic relations) and $(S_{x}^{1})^{2} + (S_{x}^{2})^{2} + (S_{x}^{3})^{2} = 2$. The low temperature phase diagram for $d\geq3$ has been investigated by several authors \cite{BO,TTI,HK2,TLMP,FKK} and it is expected to split in four regions with ferromagnetic; spin nematic; antiferromagnetic; and staggered nematic phases. They are separated by lines where the model has the larger SU(3) symmetry. See Fig. \ref{fig phd} for an illustration. We investigate it using product states in Section \ref{sec phd}.

\bfig
\centerline{\includegraphics[width=70mm]{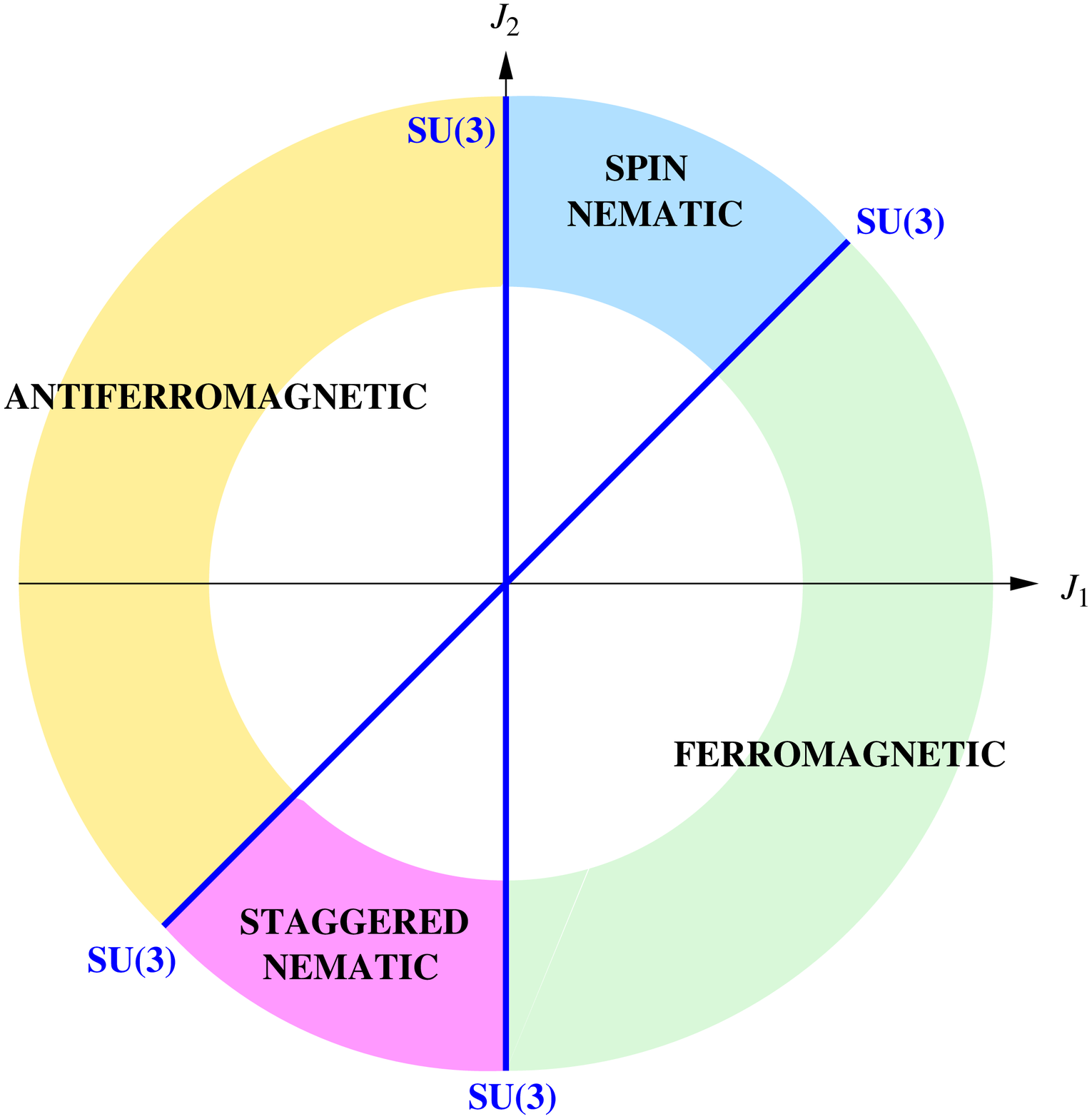}}
\caption{(Color online) Phase diagram of the spin 1 quantum model with SU(2)-invariant interactions. There are four phases, separated by the lines $J_{1}=J_{2}$ and $J_{1}=0$ where the model has the higher SU(3) symmetry.}
\label{fig phd}
\efig

There exist mathematically rigorous results that provide a few solid pillars of understanding. The presence of N\'eel order has been proved for $J_{1}<0, J_{2}=0$ and $d\geq3$ by Dyson, Lieb, and Simon \cite{DLS} using the method of infrared bounds and reflection positivity proposed by Fr\"ohlich, Simon, and Spencer \cite{FSS}. This result can be straightforwardly extended to small $J_{2}>0$. Quadrupolar order is proved in the spin nematic region for $0 \leq J_{1} \leq \frac12 J_{2}$ and $d\geq5$. For $J_{1}$ close to $\frac12 J_{2}$, the result also holds for $d\geq3$ \cite{HK2,U}. The system has actually N\'eel order in the direction $J_{1}=0, J_{2}>0$ \cite{U}. Occurrence of quadrupolar order has recently been proved for $J_{2}>0$ and small negative $J_{1}$ in $d\geq6$ \cite{Lees}; the system should have the stronger N\'eel order, though.

The goal of this article is to write down the extremal equilibrium states explicitly, and to provide evidence that all (periodic) extremal states have been identified. The method is {\it exact} in the thermodynamic limit but it is {\it not rigorous}, as several steps are postulated rather than proved mathematically. We restrict ourselves to $J_{2}\geq0$ where the model possesses probabilistic representations; quantum spin correlations are given by certain random loop correlations. In dimensions three and higher, it has been recently understood that the joint distribution of the lengths of long loops takes a universally simple form, the Poisson-Dirichlet distribution \cite{GUW}, and this allows to calculate the expectation of certain long-range two-point functions. They can also be calculated using identities from symmetry breaking. These two methods are quite independent and since they must give the same result, they provide a non-trivial test of our understanding.

This strategy has been successfully employed for spin $\frac12$ quantum Heisenberg and XY models \cite{U} and in loop representations of O($n$) models \cite{NCSOS}. There is in fact little doubt regarding the nature of extremal states and of symmetry breaking in these models. A very interesting aspect of the study of Nahum, Chalker, et.al.\ \cite{NCSOS} is to give expressions for the moments of long loops and therefore to confirm the presence of the Poisson-Dirichlet distribution.

The phase diagram is worked out in Section \ref{sec phd} using product states.
The results of the present article are then explained in Section \ref{sec extremal states}. The random loop representations are described in Sections \ref{sec simple loops} and \ref{sec loops} and the heuristics behind the joint distribution of the lengths of long loops can be found in Section \ref{sec loop lengths}. It is found that the nature of ferromagnetic and antiferromagnetic phases is as expected. The nature of SU(3) symmetry breaking is less direct, but in retrospect it is also quite expected. The quantum nematic phase, on the other hand, turns out to be surprising; this is explained in Subsection \ref{sec nematic}.

\section{Phase diagram}
\label{sec phd}

We study the nearest-neighbor interaction in order to get some understanding of the phase diagram. Let $P_{xy}^{(J)}$, $J=0,1,2$, denote the projector onto the eigensubspace of $(\vec S_{x} + \vec S_{y})^{2}$ with eigenvalue $J(J+1)$ (recall that ${\rm rank} \, P_{xy}^{(J)} = 2J+1$). It can be checked that
\be
J_{1} \vec S_{x} \cdot \vec S_{y} + J_{2} (\vec S_{x} \cdot \vec S_{y})^{2} = (3J_{2}-J_{1}) P_{xy}^{(0)} + 2J_{1} P_{xy}^{(2)} - J_{1} + J_{2}.
\ee
But many eigenstates are strongly entangled, so they have little relevance for lattices with more than two sites. A better approach is to consider product states. For this purpose, we write the interaction as a linear combination of $P_{xy}^{(0)}$ and of the transposition operator $T_{xy}$. The transposition operator is defined by
\be
\label{transposition op}
T_{xy} |a,b\rangle = |b,a\rangle,
\ee
where $a,b = -1,0,1$. As for the projector $P_{xy}^{(0)}$, its matrix elements are the Clebsch-Gordon coefficients
\be
\label{coeff P}
\langle a,b| P_{xy}^{(0)} |c,d \rangle = \tfrac13 (-1)^{a-c} \delta_{a,-b} \delta_{c,-d},
\ee
where $a,b,c,d = -1,0,1$. These operators can be expressed in terms of spin operators, namely,
\be
\begin{split}
\label{TP and S}
T_{xy} &= \vec S_{x} \cdot \vec S_{y} + (\vec S_{x} \cdot \vec S_{y})^{2} - 1, \\
P_{xy}^{(0)} &= \tfrac13 (\vec S_{x} \cdot \vec S_{y})^{2} - \tfrac13.
\end{split}
\ee
The interaction then takes the form
\be
J_{1} \vec S_{x} \cdot \vec S_{y} + J_{2} (\vec S_{x} \cdot \vec S_{y})^{2} = J_{1} T_{xy} + 3(J_{2}-J_{1}) P_{xy}^{(0)} + J_{2}.
\ee

Let $|\varphi,\psi\rangle = |\varphi\rangle \otimes |\psi\rangle$ be a product state, with $\varphi,\psi \in \bbC^{3}$. We have
\be
\label{def T}
\langle\varphi,\psi| T_{xy} |\varphi,\psi\rangle = \bigl| \langle \varphi | \psi \rangle \bigr|^{2}.
\ee
Next, let $\tilde\psi$ denote the vector in $\bbC^{3}$ with coefficients
\be
\tilde\psi_{a} = (-1)^{a} \bar\psi_{-a},
\ee
where $a = -1,0,1$. Then
\be
\label{def P}
\langle\varphi,\psi| P_{xy}^{(0)} |\varphi,\psi\rangle = \tfrac13 \bigl| \langle \varphi | \tilde\psi \rangle \bigr|^{2}.
\ee

Eqs \eqref{def T} and \eqref{def P} allow to find the product states that minimize the interaction $-J_{1} \vec S_{x} \cdot \vec S_{y} - J_{2} (\vec S_{x} \cdot \vec S_{y})^{2}$. There are four cases.
\begin{itemize}
\item Case $J_{1}>0$, $J_{2}-J_{1}<0$: We have $\langle\varphi | \psi\rangle = 1$ and $\langle\varphi | \tilde\psi\rangle = 0$. Then $\varphi=\psi$, and the coefficients of $\varphi$ satisfy
\be
\varphi_{0}^{2} - 2 \varphi_{1} \varphi_{-1} = 0.
\ee
We get in particular the ferromagnetic states such as $|1\rangle$ and $|-1\rangle$.

\item Case $J_{1}>0$, $J_{2}-J_{1}>0$: We have $\langle\varphi | \psi\rangle = 1$ and $\langle\varphi | \tilde\psi\rangle = \e{\ii\theta}$ with $\theta \in [0,2\pi]$. Then $\varphi = \psi = \e{\ii\theta} \tilde\psi$. Explicitly, the vectors must satisfy the conditions
\be
\varphi_{\pm1} = -\e{\ii\theta} \bar\varphi_{\mp1}, \quad \varphi_{0} = \e{\ii\theta} \bar\varphi_{0}.
\ee
This gives the ``classical'' nematic state $|1\rangle + |-1\rangle$, but also $|0\rangle$ and such linear combination as $|1\rangle + \ii |0\rangle + |-1\rangle$.

\item Case $J_{1}<0$, $J_{2}-J_{1}>0$: This is identical to the first case (the ferromagnetic one) if we exchange $\psi$ with $\tilde\psi$. We get the same states for $\varphi$ but the resulting tensor state is $\varphi \otimes \tilde\varphi$. This gives in particular the N\'eel state $|1\rangle \otimes |-1\rangle$.

\item Case $J_{1}<0$, $J_{2}-J_{1}<0$: We have $\langle\varphi | \psi\rangle = \langle\varphi | \tilde\psi\rangle = 0$. There are many solutions; we can choose a nematic state that satisfies $\varphi = \e{\ii\theta} \tilde\varphi$, and any state $\psi$ that is orthogonal to $\varphi$.
\end{itemize}

This analysis suggests the phase diagram depicted in Fig.\ \ref{fig phd}.

\section{Nature of extremal states}
\label{sec extremal states}

Let $\langle \cdot \rangle_{H_{\Lambda}}$ denote the finite-volume Gibbs state with Hamiltonian $H_{\Lambda}$, where the expectation of the observable $A$ is
\be
\langle A \rangle_{H_{\Lambda}} = \frac1{Z_{\Lambda}} \Tr A \e{-\beta H_{\Lambda}},
\ee
with $Z_{\Lambda} = \Tr \e{-\beta H_{\Lambda}}$ the partition function. We let $\langle \cdot \rangle$ denote its infinite-volume limit. Our goal is to identify the extremal states $\langle \cdot \rangle_{\vec a}$ such that
\be
\langle \cdot \rangle = \int \langle \cdot \rangle_{\vec a} \;\dd\nu(\vec a),
\ee
where $\dd\nu$ is a probability measure on the set of possible values for $\vec a$. We describe separately the ferromagnetic, broken SU(3), spin nematic, broken staggered SU(3), and antiferromagnetic phases. In each case we propose a decomposition, use it to calculate the two-point functions, and compare with results obtained with random loop representations. We conclude the section with a discussion of the staggered nematic phase; we make a conjecture but do not provide conclusive evidence.

\subsection{The ferromagnetic phase}

This is the simplest situation, valid when parameters satisfy $J_{1}>0$ and $J_{2}<J_{1}$ (see Fig.\ \ref{fig phd}). To each unit vector $\vec a \in \bbR^{3}$, there corresponds the ferromagnetic extremal state
\be
\langle \cdot \rangle_{\vec a}^{\rm ferro} = \lim_{h\to0+} \lim_{\Lambda\to\bbZ^{d}} \langle \cdot \rangle_{H_{\Lambda} - h \sum_{x} \vec a \cdot \vec S_{x}}.
\ee
Let $\bbS^{2}$ denote the unit sphere in $\bbR^{3}$ and let $\dd\nu$ be the uniform probability measure. The infinite-volume state $\langle \cdot \rangle$ should satisfy the identity
\be
\label{ferro decomposition}
\langle \cdot \rangle = \int_{\bbS^{2}} \langle \cdot \rangle^{\rm ferro}_{\vec a} \;\dd\nu(\vec a).
\ee
We can apply this decomposition to the special case of the two-point function $\langle S_{x}^{3} S_{y}^{3} \rangle$ with $x,y$ far apart. We have
\be
\label{calcul ferro}
\begin{split}
\langle S_{x}^{3} S_{y}^{3} \rangle &= \tfrac13 \langle \vec S_{x} \cdot \vec S_{y} \rangle \equal{(a)} \tfrac13 \int_{\bbS^{2}} \langle \vec S_{x} \cdot \vec S_{y} \rangle^{\rm ferro}_{\vec a} \;\dd\nu(\vec a) \\
&\equal{(b)} \tfrac13 \langle \vec S_{x} \cdot \vec S_{y} \rangle^{\rm ferro}_{\vec e_{3}} \equal{(c)} \tfrac13 \sum_{i=1}^{3} \langle S_{x}^{i} \rangle^{\rm ferro}_{\vec e_{3}} \langle S_{y}^{i} \rangle^{\rm ferro}_{\vec e_{3}} \\
&\equal{(d)} \tfrac13 \Bigl( \langle S_{x}^{3} \rangle^{\rm ferro}_{\vec e_{3}} \Bigr)^{2}.
\end{split}
\ee
We have used
\begin{itemize}
\item[(a)] the conjecture Eq.\ \eqref{ferro decomposition};
\item[(b)] $\vec S_{x} \cdot \vec S_{y}$ is SU(2)-invariant;
\item[(c)] extremal states are clustering, hence factorization when $\|x-y\| \gg 1$;
\item[(d)] translation invariance and $\langle S_{x}^{i} \rangle^{\rm ferro}_{\vec e_{3}} = 0$ for $i=1,2$; the latter follows from the rotation invariance around $\vec e_{3}$ of the extremal state.
\end{itemize}

We provide independent evidence for the identity \eqref{calcul ferro}, and hence \eqref{ferro decomposition}, using the random loop representation and the conjecture about the joint distribution of the lengths of long loops; see Subsection \ref{sec justifications}.

\subsection{Breaking SU(3) invariance}
\label{sec SU(3)}

When $J_{1}=J_{2}$ the Hamiltonian has the larger SU(3) symmetry. Indeed, the interaction is given by the transposition operator $T_{xy}$ defined in \eqref{transposition op}. Given any unitary $U$ in $\bbC^{3}$, the tensor product $U_{\Lambda} = \otimes_{x} U$ commutes with all $T_{xy}$'s.

The group SU(3) has eight dimensions and it is generated e.g.\ by the Gell-Mann matrices $\lambda^{1},\dots,\lambda^{8}$. All matrices have empty diagonal except $\lambda^{3}$ and $\lambda^{8}$:
\be
\lambda^{3} = \Bigl( \begin{smallmatrix} 1 & 0 & 0 \\ 0 & -1 & 0 \\ 0 & 0 & 0 \end{smallmatrix} \Bigr), \qquad \lambda^{8} = \tfrac1{\sqrt3} \Bigl( \begin{smallmatrix} 1 & 0 & 0 \\ 0 & 1 & 0 \\ 0 & 0 & -2 \end{smallmatrix} \Bigr).
\ee
Another property that is relevant for our purpose is the identity
\be
\label{GM vs T}
\sum_{i=1}^{8} \lambda_{x}^{i} \lambda_{y}^{i} = 2 T_{xy} - \tfrac23,
\ee
which shows that $\sum_{i} \lambda_{x}^{i} \lambda_{y}^{i}$ is SU(3)-invariant. The extremal states are given by
\be
\langle \cdot \rangle_{\vec a} = \lim_{h\to0+} \lim_{\Lambda \to \bbZ^{d}} \langle \cdot \rangle_{H_{\Lambda} - h\sum_{x} \vec a \cdot \vec\lambda_{x}}.
\ee
Here, $\vec a$ is a vector in $\bbS^{7}$, the unit sphere in $\bbR^{8}$. Let $\dd\nu$ be the uniform probability measure on $\bbS^{7}$. The decomposition of the SU(3)-invariant Gibbs state is
\be
\label{decomposition SU(3)}
\langle \cdot \rangle = \int_{\bbS^{7}} \langle \cdot \rangle_{\vec a} \;\dd\nu(\vec a).
\ee
The two-point function can be calculated as in the ferromagnetic case, Eqs \eqref{calcul ferro}:
\be
\label{calcul SU(3)}
\begin{split}
\langle S_{x}^{3} S_{y}^{3} \rangle &= \tfrac18 \Big\langle \sum_{i=1}^{8} \lambda_{x}^{i} \lambda_{y}^{i} \Big\rangle = \tfrac18 \int_{\bbS^{7}} \Big\langle \sum_{i=1}^{8} \lambda_{x}^{i} \lambda_{y}^{i} \Big\rangle_{\vec a} \;\dd\nu(\vec a) \\
&= \tfrac18 \Big\langle \sum_{i=1}^{8} \lambda_{x}^{i} \lambda_{y}^{i} \Big\rangle_{\vec e_{3}} = \tfrac18 \sum_{i=1}^{8} \bigl( \langle \lambda_{x}^{i} \rangle_{\vec e_{3}} \bigr)^{2} \\
&= \tfrac18 \bigl( \langle \lambda_{x}^{3} \rangle_{\vec e_{3}} \bigr)^{2} + \tfrac18 \bigl( \langle \lambda_{x}^{8} \rangle_{\vec e_{3}} \bigr)^{2}.
\end{split}
\ee
We used the clustering property of extremal states when $\|x-y\|\to\infty$.
We have $\langle \lambda_{x}^{i} \rangle_{\vec e_{3}} = 0$ if $i \neq 3,8$, and $\langle \lambda_{x}^{8} \rangle_{\vec e_{3}} = \frac1{\sqrt3} \langle \lambda_{x}^{3} \rangle_{\vec e_{3}}$. We obtain
\be
\label{equality SU(3)}
\langle S_{x}^{3} S_{y}^{3} \rangle = \tfrac16 \bigl( \langle \lambda_{x}^{3} \rangle_{\vec e_{3}} \bigr)^{2}.
\ee
We check this identity using the random loop representation in Subsection \ref{sec justifications}.

The matrix $\lambda^{8}$ is peculiar, as it is not unitarily equivalent to the other seven matrices. It is worth checking that the corresponding extremal state gives the same result. The second line of Eq.\ \eqref{calcul SU(3)} becomes
\be
\begin{split}
\dots &= \tfrac18 \Big\langle \sum_{i=1}^{8} \lambda_{x}^{i} \lambda_{y}^{i} \Big\rangle_{\vec e_{8}} = \tfrac18 \sum_{i=1}^{8} \bigl( \langle \lambda_{x}^{i} \rangle_{\vec e_{8}} \bigr)^{2} \\
&= \tfrac18 \bigl( \langle \lambda_{x}^{1} \rangle_{\vec e_{8}} \bigr)^{2} + \tfrac18 \bigl( \langle \lambda_{x}^{8} \rangle_{\vec e_{8}} \bigr)^{2}.
\end{split}
\ee
It turns out that $\langle \lambda_{x}^{8} \rangle_{\vec e_{8}} = \frac1{\sqrt3} \langle \lambda_{x}^{1} \rangle_{\vec e_{8}} = \frac1{\sqrt3} \langle \lambda_{x}^{3} \rangle_{\vec e_{3}}$. This can indeed be seen using the random loop representation. We recover the identity \eqref{equality SU(3)}.

\subsection{The curious spin nematic phase}
\label{sec nematic}

In the classical model with $J_{1}=0, J_{2}>0$, there is a spin nematic phase where the classical spins are aligned or anti-aligned \cite{AZ,BC}. This suggests that extremal states can be defined using an external field of the form $-\sum_{x} (\vec a \cdot \vec S_{x})^{2}$. The quantum system is very different for two main reasons. First, the direction $J_{1}=0, J_{2}>0$ is not nematic but it has N\'eel order \cite{U}; its broken staggered SU(3) invariance is explained in the next subsection. The second reason is that for $J_{2} > J_{1} > 0$, extremal states are obtained using the external field $+\sum_{x} (\vec a \cdot \vec S_{x})^{2}$ with a ``$+$'' sign. Rather than projecting onto the eigenstates of $S_{x}^{i}$ with eigenvalues $\pm1$, we project onto the eigensubspace with value 0! This is rather surprising and this signals a purely quantum phenomenon.

Let $\bbS^{2}_{+}$ denote the unit hemisphere in $\bbR^{3}$ with nonnegative third coordinates. Given $\vec a \in \bbS^{2}_{+}$, the corresponding extremal state is
\be
\label{extremal nem state}
\langle \cdot \rangle_{\vec a}^{\rm nem} = \lim_{h\to0+} \lim_{\Lambda \to \bbZ^{d}} \langle \cdot \rangle_{H_{\Lambda} + h \sum_{x} (\vec a \cdot \vec S_{x})^{2}}.
\ee
With $\dd\nu$ the uniform probability measure on $\bbS^{2}_{+}$, the nematic decomposition reads
\be
\label{decomposition nem}
\langle \cdot \rangle = \int_{\bbS^{2}_{+}} \langle \cdot \rangle_{\vec a}^{\rm nem} \;\dd\nu(\vec a).
\ee

The spin-spin correlation function is not a suitable order parameter for the nematic phase. Indeed, it certainly has exponential decay, as suggested by its random loop representation; see Eq.\ \eqref{interesting spin-spin} below. So we rather pick the observable for quadrupolar order:
\be
A_{x} = (S_{x}^{3})^{2} - \tfrac23.
\ee
Notice that $\langle A_{x} \rangle = 0$. The calculation of the quadrupolar two-point function can be done in the nematic phase as follows.
\be
\begin{split}
\langle A_{x} A_{y} \rangle &= \int_{\bbS^{2}_{+}} \langle A_{x} A_{y} \rangle_{\vec a}^{\rm nem} \;\dd\nu(\vec a) \\
& = \int_{\bbS^{2}_{+}} \Bigl( \big\langle (S_{x}^{3})^{2} - \tfrac23 \big\rangle_{\vec a}^{\rm nem} \Bigr)^{2} \;\dd\nu(\vec a) \\
&= \int_{\bbS^{2}_{+}} \Bigl( \big\langle (\vec a \cdot \vec S_{x})^{2} - \tfrac23 \big\rangle_{\vec e_{3}}^{\rm nem} \Bigr)^{2} \;\dd\nu(\vec a).
\end{split}
\ee
The second identity is due to clustering when $\|x-y\| \gg 1$ and the last identity follows by rotating the observable rather than the state. Next, we have
\be
\begin{split}
\big\langle (\vec a \cdot \vec S_{x})^{2} - \tfrac23 \big\rangle_{\vec e_{3}}^{\rm nem} &= \sum_{i,j=1}^{3} a_{i} a_{j} \; \langle S_{x}^{i} S_{x}^{j} \rangle_{\vec e_{3}}^{\rm nem} - \tfrac23 \\
&= \sum_{i=1}^{3} a_{i}^{2} \; \big\langle (S_{x}^{i})^{2} - \tfrac23 \big\rangle_{\vec e_{3}}^{\rm nem}.
\end{split}
\ee
We now use the following identities:
\be
\big\langle (S_{x}^{3})^{2} - \tfrac23 \big\rangle_{\vec e_{3}}^{\rm nem} = -2 \big\langle (S_{x}^{1})^{2} - \tfrac23 \big\rangle_{\vec e_{3}}^{\rm nem} = -2 \big\langle (S_{x}^{2})^{2} - \tfrac23 \big\rangle_{\vec e_{3}}^{\rm nem}.
\ee
They follow from symmetry and from the identity $(S_{x}^{1})^{2} + (S_{x}^{2})^{2} + (S_{x}^{3})^{2} = 2$. We obtain
\be
\label{calcul nem}
\begin{split}
\langle A_{x} A_{y} \rangle &= \Bigl( \big\langle (S_{x}^{1})^{2} - \tfrac23 \big\rangle_{\vec e_{3}}^{\rm nem} \Bigr)^{2} \int_{\bbS_{+}^{2}} (a_{1}^{2} + a_{2}^{2} - 2a_{3}^{2})^{2} \dd\nu(\vec a) \\
&= \tfrac45 \Bigl( \big\langle (S_{x}^{1})^{2} - \tfrac23 \big\rangle_{\vec e_{3}}^{\rm nem} \Bigr)^{2}.
\end{split}
\ee
This identity is verified in Subsection \ref{sec justifications} using the loop representation.

It is natural to wonder about the states $\langle \cdot \rangle_{\vec a}^{*}$ obtained with external field $-\sum_{x} (\vec a \cdot \vec S_{x})^{2}$. If these were the extremal states, one should expect the decomposition $\langle \cdot \rangle = \int_{\bbS_{+}^{2}} \langle \cdot \rangle_{\vec a}^{*} \;\dd\nu(\vec a)$. However, the calculations of $\langle A_{x} A_{y} \rangle$ give a result that is incompatible with the random loop calculations, so this has to be discarded. If $\langle \cdot \rangle_{\vec a}^{*}$ are not extremal states, then they should be linear combinations of the nematic states. And indeed, it appears that
\be
\label{faux etats purs}
\langle \cdot \rangle_{\vec a}^{*} = \frac1{2\pi} \int_{\vec b \perp \vec a} \langle \cdot \rangle_{\vec b}^{\rm nem} \, \dd\vec b.
\ee
In the case of one-site observables this can be checked using the following identity, that holds for all matrices $A$ in $\bbC^{3}$:
\be
\label{decomposition stag SU(3)}
\frac1{2\pi} \int_{\vec b \perp \vec e_{3}} \langle 0 | \, U_{\vec b} \, A \, U_{\vec b}^{*} \, |0\rangle \;\dd\vec b = \tfrac12 \Tr (S^{3})^{2} A.
\ee
Here, we set $U_{\vec b} = \e{\ii \frac\pi2 (\vec b \times \vec e_{3}) \cdot \vec S}$. The identity \eqref{faux etats purs} can certainly be verified for more general many-site observables.

\subsection{Breaking staggered SU(3) invariance}

In the case $J_{1}=0$, $J_{2}>0$, the interaction is equivalent to the projector onto the spin singlet $P_{xy}^{(0)}$ defined in \eqref{coeff P}.
Given any unitary matrix $U$ in $\bbC^{3}$, let $\widetilde U$ denote the matrix with elements
\be
\widetilde U_{a,b} = (-1)^{a-b} \overline{U}_{-a,-b}.
\ee
Here, $\overline{U}_{-a,-b}$ is the complex conjugate of $U_{-a,-b}$; notice that $\widetilde{\widetilde U} = U$.
One can check that
\be
P_{xy}^{(0)} \; U \otimes \widetilde U = U \otimes \widetilde U \; P_{xy}^{(0)} = P_{xy}^{(0)}.
\ee
As a consequence, the Hamiltonian $H_{\Lambda}$ has SU(3) invariance whenever the lattice $\Lambda$ is bipartite. More precisely, if $\Lambda_{\rm A}, \Lambda_{\rm B}$ are the two sublattices, the unitary operator on $\otimes_{x\in\Lambda} \bbC^{3}$ is
\be
U_{\Lambda} = \otimes_{x \in \Lambda_{\rm A}} U \; \otimes_{x \in \Lambda_{\rm B}} \widetilde U.
\ee

On non-bipartite lattices, the Hamiltonian has only SU(2) invariance and the low temperature phase is the spin nematic phase discussed in the previous subsection.

In the rest of this subsection we restrict to the bipartite lattice $\bbZ^{d}$, $d\geq3$. The extremal states can be defined using staggered fields. Given a unit vector $\vec a$ in $\bbR^{8}$, let
\be
Q_{\Lambda}^{\vec a} = \sum_{x\in\Lambda_{\rm A}} \vec a \cdot \vec\lambda_{x} + \sum_{x\in\Lambda_{\rm B}} \vec a \cdot \vec{\widetilde\lambda}_{x}.
\ee
As in Subsection \ref{sec SU(3)}, $\vec\lambda_{x} = (\lambda_{x}^{1}, \dots, \lambda_{x}^{8})$ is the vector of Gell-Mann matrices acting on the site $x$. We also defined $\vec{\widetilde\lambda}_{x} = (\widetilde\lambda_{x}^{1}, \dots, \widetilde\lambda_{x}^{8})$. Then we set
\be
\langle \cdot \rangle_{\vec a} = \lim_{h\to0+} \lim_{\Lambda\to\bbZ^{d}} \langle \cdot \rangle_{H_{\Lambda} - h Q_{\Lambda}^{\vec a}}.
\ee
The decomposition of the symmetric Gibbs state reads
\be
\langle \cdot \rangle = \int_{\bbS^{7}} \langle \cdot \rangle_{\vec a} \;\dd\nu(\vec a).
\ee
The justification of this formula mirrors that of Eq.\ \eqref{decomposition SU(3)} in the non-staggered SU(3) situation. Without entering details, let us note that
\be
3 P_{xy}^{(0)} = \left( \begin{smallmatrix}
0 & 0 & 0 & 0 & 0 & 0 & 0 & 0 & 0 \\
0 & 0 & 0 & 0 & 0 & 0 & 0 & 0 & 0 \\
0 & 0 & 1 & 0 & -1 & 0 & 1 & 0 & 0 \\
0 & 0 & 0 & 0 & 0 & 0 & 0 & 0 & 0 \\
0 & 0 & -1 & 0 & 1 & 0 & -1 & 0 & 0 \\
0 & 0 & 0 & 0 & 0 & 0 & 0 & 0 & 0 \\
0 & 0 & 1 & 0 & -1 & 0 & 1 & 0 & 0 \\
0 & 0 & 0 & 0 & 0 & 0 & 0 & 0 & 0 \\
0 & 0 & 0 & 0 & 0 & 0 & 0 & 0 & 0
\end{smallmatrix} \right)
= \tfrac12 \sum_{i=1}^{8} \lambda^{i} \otimes \widetilde\lambda^{i} + \tfrac13.
\ee
It follows that $\vec\lambda_{x} \cdot \vec{\widetilde\lambda}_{y}$ is left invariant by $U_{\Lambda}$ whenever $x \in \Lambda_{\rm A}, y \in \Lambda_{\rm B}$. From Eq.\ \eqref{GM vs T}, $\vec\lambda_{x} \cdot \vec\lambda_{y}$ is also left invariant whenever $x,y$ belong to the same sublattice. The loop representation differs from the one for $J_{1}=J_{2}$ but the joint distribution of macroscopic loops is the same PD(3).

\subsection{The antiferromagnetic phase}

The extremal states are indexed by unit vectors $\vec a$ in $\bbR^{3}$. If the lattice is bipartite they can be defined using the usual staggered fields, namely
\be
\langle \cdot \rangle_{\vec a}^{\rm AF} = \lim_{h\to0+} \lim_{\Lambda\to\bbZ^{d}} \langle \cdot \rangle_{H_{\Lambda} - h \sum_{x\in\Lambda} (-1)^{x} \vec a \cdot \vec S_{x}}.
\ee
The decomposition reads
\be
\label{decomposition AF}
\langle \cdot \rangle = \int_{\bbS^{2}} \langle \cdot \rangle_{\vec a}^{\rm AF} \;\dd\nu(\vec a).
\ee
The justification for this formula is quite identical to that of \eqref{ferro decomposition} in the ferromagnetic regime. Loops are different but the joint distribution of macroscopic loops is the same PD(2).

Notice that, for $J_{1}<0$, the random loop representation is only valid when the lattice is bipartite. There are problematic signs otherwise. The study of frustrated systems is notoriously difficult.

\subsection{The staggered nematic phase}

We conclude this section with a conjecture about the likely extremal states in the region $J_{2} < J_{1} <0$. The study of product states in Section \ref{sec phd} suggests that each edge consists of a nematic state, and of a state that is perpendicular with respect to the scalar product in $\bbC^{3}$. On a bipartite lattice one gains entropy by choosing a given nematic state on all sites of a sublattice; this allows to choose independent perpendicular states on the sites of the other sublattice. We saw in Subsection \ref{sec nematic} that $|0\rangle$ seems better. This suggests the following extremal states; if $\vec a \in \bbS^{2}_{+}$,
\be
\label{conj st nem}
\langle \cdot \rangle_{\vec a}^{\text{st.\ nem.}} = \lim_{h\to0+} \lim_{\Lambda\to\bbZ^{d}} \langle \cdot \rangle_{H_{\Lambda} + h\sum_{x} (-1)^{x} (\vec a \cdot \vec S)^{2}}.
\ee
And if $\vec a \in \bbS^{2}_{-}$, $\langle \cdot \rangle_{\vec a}^{\text{st.\ nem.}}$ is as above but with $h\to0-$. In the first case, nematic order occurs on the sublattice that contains the origin; in the second case, order occurs on the other sublattice.

There exist random loop representations for these parameters, but they carry signs and therefore do not offer evidence for the conjecture \eqref{conj st nem}.

\section{Random loop representation for the nematic and SU(3) regions}
\label{sec simple loops}

The partition function of quantum lattice models can be expanded using the Trotter product formula, or the Duhamel formula, yielding a sort of classical model in one more dimensions. For some models the expansion takes the form of random loops. This turns out to be important for our purpose.

Random loop models were used in studies of the spin $\frac12$ Heisenberg ferromagnet \cite{Pow,Toth} and of the antiferromagnet \cite{AN}. An extension was proposed recently \cite{U} that applies to the spin $\frac12$ XY model, and to the spin 1 model in the nematic and SU(3) regions $0 \leq J_{1} \leq J_{2}$. We describe it in details in this section. A more complicated class of loop models that applies to the half-plane $J_{2}\geq0$ will be introduced in the next section.

Recall the transposition operator $T_{xy}$ in Eq.\ \eqref{transposition op}, the projector onto the spin singlet $P_{xy}^{(0)}$ in \eqref{coeff P}, and their relations with spin operators in \eqref{TP and S}. We consider convex combinations of these operators. Namely, we consider the family of Hamiltonians
\be
\begin{split}
H_{\Lambda} &= -\sum_{\langle x,y \rangle} \bigl( u T_{xy} + (1-u) P_{xy}^{(0)} -1 \bigr) \\
&= -\sum_{\langle x,y \rangle} \bigl( u \vec S_{x} \cdot \vec S_{y} + (\vec S_{x} \cdot \vec S_{y})^{2} -2 \bigr).
\end{split}
\ee

Next, recall that a Poisson point process on the interval $[0,1]$ describes the occurrence of independent events at random times. Let $u\geq0$ the intensity of the process. The probability that an event occurs in the infinitesimal interval $[t,t+\dd t]$ is $u \dd t$; disjoint intervals are independent. Poisson point processes are relevant to us because of the following expansion of the exponential of matrices:
\be
\exp\Bigl\{ u \sum_{i=1}^{k} (M_{i}-1) \Bigr\} = \int \rho(\dd\omega) \prod_{(i,t) \in \omega} M_{i},
\ee
where $\rho$ is a Poisson point process on $\{1,\dots,k\} \times [0,1]$ with intensity $u$, and the product is over the events of the realization $\omega$ in increasing times. We actually consider an extension where the time intervals are labeled by the edges of the lattice, and where two kinds of events occur with respective intensities $u$ and $1-u$. Then
\bm
\label{Poisson exp}
\exp\Bigl\{ -\sum_{\langle x,y \rangle} \bigl( uM_{xy}^{(1)} + (1-u) M_{xy}^{(2)} - 1 \bigr) \Bigr\} \\ = \int\rho(\dd\omega) \prod_{(x,y,i,t) \in \omega} M_{xy}^{(i)}.
\end{multline}
The product is over the events of $\omega$ in increasing times; the label $i$ is equal to 1 if the event is of the first kind, and 2 if the event is of the second kind.

Let $|\sigma\rangle = \otimes_{x\in\Lambda} |\sigma_{x}\rangle$, $\sigma_{x} \in \{-1,0,1\}$, denote the elements of the basis of $\caH_{\Lambda}$ where $S_{x}^{3}$ are diagonal. Applying the Poisson expansion \eqref{Poisson exp}, we get
\bm
\label{full expansion}
\Tr \e{-\sum_{\langle x,y \rangle} (uT_{xy} + (1-u) P_{xy}^{(0)} -1)} = \int\rho(\dd\omega) \sum_{|\sigma_{1}\rangle, \dots, |\sigma_{k}\rangle} \\
\langle\sigma_{1}| M_{x_{k} y_{k}}^{(i_{k})} |\sigma_{k}\rangle \langle\sigma_{k}| M_{x_{k-1} y_{k-1}}^{(i_{k-1})} |\sigma_{k-1}\rangle \dots \langle\sigma_{2}| M_{x_{1} y_{1}}^{(i_{1})} |\sigma_{1}\rangle.
\end{multline}
Here, $(x_{1}, y_{1}, i_{1}), \dots, (x_{k},y_{k},i_{k})$ are the events of the realization $\omega$ in increasing times. Their number $k$ is random.

This expansion has a convenient graphical description. Namely, let $\dd\rho$ denote the measure for a Poisson point process for each edge of $\Lambda$, where ``crosses'' occur with intensity $u$ and ``double bars'' occur with intensity $1-u$. In order to find the loop that contains a given point $(x,t) \in \Lambda \times [0,\beta]$, one can start by moving upwards, say, until one meets a cross or a double bar. Then one jumps onto the corresponding neighbor; if the transition is a cross, one continues in the same vertical direction; if it is a double bar, one continues in the opposite direction. The vertical direction has periodic boundary conditions. See Fig.\ \ref{fig loops} for illustrations.

\bfig
\centerline{\includegraphics[width=85mm]{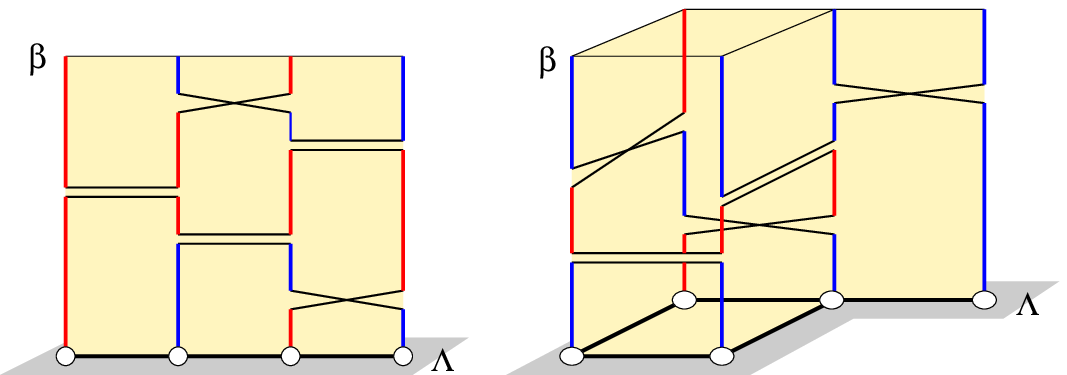}}
\caption{(Color online) The random loop model for parameters $0 \leq J_{1} \leq J_{2}$. In both figures, the realization has $|\caL(\omega)|=2$ loops.}
\label{fig loops}
\efig

The sum over $|\sigma_{i}\rangle$ is then equivalent to assigning independent labels along each loop. The labels take values in $\{-1,0,1\}$ and they are constant along vertical legs. When meeting a cross and jumping onto a neighbor, they stay constant. When meeting a double bar and jumping onto a neighbor, they change sign. In \eqref{full expansion} the matrix elements of $T_{xy}$ are equal to 1 if the labels are compatible with the loop and 0 otherwise. The matrix elements of $P_{xy}^{(0)}$ are $\pm1$ if the labels are compatible with the loop and 0 otherwise. The number of factors $-1$ along a given loop is always even, because it is related to the number of changes of vertical directions. Thus the overall factor is 1 and the signs disappear.

Since each loop can have exactly three labels, the partition function is
\be
Z_{\Lambda} = \Tr_{\caH_{\Lambda}} \e{-\beta H_{\Lambda}} = \int 3^{|\caL(\omega)|} \rho(\dd\omega).
\ee
Let $\bbP$ denote the probability with respect to the measure $\frac1{Z_{\Lambda}} 3^{|\caL(\omega)|} \rho(\dd\omega)$. The spin-spin correlation function can be calculated using the same expansion as for the partition function. We get
\be
\Tr S_{x}^{3} S_{y}^{3} \e{-\beta H_{\Lambda}} = \int\rho(\dd\omega) \sum_{\sigma:\omega} \sigma_{x} \sigma_{y}.
\ee
The sum is over all possible labels for the loops, and $\sigma_{x}$ denotes the label at site $x$ and time 0. The sum is zero unless $x$ and $y$ belong to the same loop (at time 0), in which case it gives $\pm \frac23 3^{|\caL(\omega)|}$. The sign depends on whether the loop reaches $y$ from below or from above. We find therefore
\be
\label{interesting spin-spin}
\begin{tikzpicture}[>=angle 90]
\node (n0) at (0,0) {$\langle S_{x}^{i} S_{y}^{i} \rangle_{H_{\Lambda}} = \tfrac23 \Bigl[ \bbP\Bigl($};
\node (n1) at (1.5,0) {$x$};
\node (n2) at (2.2,-0.04) {$y$};
\draw[-] (n1) edge[out=90,in=270] (n2);
\draw[-] (n1) edge[out=270,in=90] (n2);
\node (n3) at (2.8,0) {$\Bigr) - \bbP \Bigl($};
\node (n4) at (3.4,0) {$x$};
\node (n5) at (4.0,-0.04) {$y$};
\draw[-] (n4) edge[out=90,in=90] (n5);
\draw[-] (n4) edge[out=270,in=270] (n5);
\node (n6) at (4.3,0) {$\Bigr) \Bigr].$};
\end{tikzpicture}
\ee
The first event refers to loops that connect the top of $x$ to the bottom of $y$; the second event refers to loops that connect the top of $x$ with the top of $y$. This expression is actually very interesting and suggests that the spin-spin correlation function has exponential decay in the nematic phase, for all temperatures. The quadrupolar correlation function can also be calculated \cite{U} and it is proportional to the probability that both sites (at time 0) belong to the same loop:
\be
\langle A_{x} A_{y} \rangle_{H_{\Lambda}} = \tfrac29 \bbP \bigl( x \leftrightarrow y \bigr).
\ee

\section{Random loop representations for $J_{2} \geq 0$}
\label{sec loops}

We now describe a more general class of loop models that apply to the upper half of the phase diagram, $J_{2}\geq0$. These loop models were proposed by Nachtergaele \cite{Nac1,Nac2} and independently by Harada and Kawashima \cite{HK1}. In the region $J_{1} < 0$, the lattice must be bipartite in order for the signs to be positive.

\subsection{Transition operators}

Consider a system where each site $x\in\Lambda$ hosts two spins $\frac12$. The Hilbert space is $\caH_{\Lambda}' = \otimes_{x\in\Lambda} (\bbC^{2} \otimes \bbC^{2})$. Let $\caS$ the projector onto the triplet subspace in $\bbC^{2} \otimes \bbC^{2}$,
\be
\label{sym}
\caS |a,b\rangle = \tfrac12 |a,b\rangle + \tfrac12 |b,a\rangle,
\ee
and let
\be
\caS_{x} = \caS \otimes \bbone_{\Lambda\setminus\{x\}}, \quad \caS_{\Lambda} = \otimes_{x\in\Lambda} \caS.
\ee
Let $V : \bbC^{3} \to \bbC^{2} \otimes \bbC^{2}$ be the operator such that
\be
\begin{split}
&V^{*} V = \bbone_{\bbC^{3}}, \\
&VV^{*} = \caS.
\end{split}
\ee
Define the operators on $\bbC^{2} \otimes \bbC^{2}$ by
\be
T^{i} = V S^{i} V^{*}, \quad i=1,2,3,
\ee
where the $S^{i}$s are the spin operators in $\bbC^{3}$. One can check that the $T^{i}$s satisfy the following relations:
\be
\begin{split}
&[T^{1},T^{2}] = \ii T^{3}, \quad \text{(and cyclic relations)} \\
&(T^{1})^{2} + (T^{2})^{2} + (T^{3})^{2} = 2\caS.
\end{split}
\ee
These operators can be written with the help of Pauli matrices. Namely,
\be
\label{T Pauli}
\begin{split}
T^{i} &= (\sigma^{i} \otimes \bbone + \bbone \otimes \sigma^{i}) \caS \\
&= 2 \caS (\sigma^{i} \otimes \bbone) \caS.
\end{split}
\ee
Notice that $\sigma^{i} \otimes \bbone + \bbone \otimes \sigma^{i}$ commutes with $\caS$.

Using the operators above, we define the following Hamiltonian on $\caH_{\Lambda}'$:
\be
H_{\Lambda}' = -\sum_{\langle x,y \rangle} \bigl( J_{1} \vec T_{x} \cdot \vec T_{y} + J_{2} (\vec T_{x} \cdot \vec T_{y})^{2} \bigr).
\ee
The partition function and the Gibbs states are defined by
\be
\begin{split}
&Z_{\Lambda}' = \Tr_{\caH_{\Lambda}'} \caS_{\Lambda} \e{-\beta H_{\Lambda}'}, \\
&\langle A' \rangle_{H_{\Lambda}'} = \frac1{Z_{\Lambda}'} \Tr A' \caS_{\Lambda} \e{-\beta H_{\Lambda}'}.
\end{split}
\ee
The correspondence between the spin 1 model $H_{\Lambda}$ and the new model $H_{\Lambda}'$ is then
\be
\label{corresp}
\langle A \rangle_{H_{\Lambda}} = \langle V_{\Lambda} A V_{\Lambda}^{*} \rangle_{H_{\Lambda}'}.
\ee

We need to rewrite the Hamiltonian $H_{\Lambda}'$ using suitable operators in order to get the loop representations. We use symbols that are suggestive of the two sites of two spins each, and of the links due to transitions. We also use the notation $|a,b\rangle$ for an element in the one-site Hilbert space $\bbC^{2} \otimes \bbC^{2}$, and $|a,b\rangle \otimes |c,d\rangle$ for an element in the two-site Hilbert space. With these conventions in mind, let
\be
\begin{split}
&\tra' \; |a,b\rangle \otimes |c,d\rangle = |c,b\rangle \otimes |a,d\rangle, \\
&\trb' \; |a,b\rangle \otimes |c,d\rangle = |c,d\rangle \otimes |a,b\rangle.
\end{split}
\ee
These are transposition operators for one or two spins. We actually need the symmetrized operators
\be
\tra = \caS_{x,y} \; \tra' \caS_{x,y}, \qquad \trb = \caS_{x,y} \; \trb' \caS_{x,y}.
\ee
They can be written in terms of spin operators as follows.
\be
\begin{split}
&\tra \; = \tfrac12 \vec T_{x} \cdot \vec T_{y} + \tfrac12 \caS_{x,y}, \\
&\trb \; = \vec T_{x} \cdot \vec T_{y} + (\vec T_{x} \cdot \vec T_{y})^{2} - \caS_{x,y}.
\end{split}
\ee
Next, let us consider the operators that are related to spin singlets of spins at distinct sites.
\be
\label{matrix elements}
\begin{split}
\langle a',b'| & \otimes \langle c',d'| \; \trc' |a,b\rangle \otimes |c,d\rangle \\
&= (-1)^{a-a'} (-1)^{b-b'} \delta_{a,-d} \delta_{b,-c} \delta_{a',-d'} \delta_{b',-c'}, \\
\langle a',b'| & \otimes \langle c',d'| \; \trd' |a,b\rangle \otimes |c,d\rangle \\
&= (-1)^{b-b'} \delta_{a,a'} \delta_{d,d'} \delta_{b,-c} \delta_{b',-c'}.
\end{split}
\ee
We again need the symmetrized operators
\be
\trc = \caS_{x,y} \; \trc' \caS_{x,y}, \qquad \trd = \caS_{x,y} \; \trd' \caS_{x,y}.
\ee
In terms of spin operators, they are given by
\be
\begin{split}
&\trc \; = (\vec T_{x} \cdot \vec T_{y})^{2} - \caS_{x,y}, \\
&\trd \; = -\tfrac12 \vec T_{x} \cdot \vec T_{y} + \tfrac12 \caS_{x,y}.
\end{split}
\ee
We can also consider the operator
\bm
\langle a',b'| \otimes \langle c',d'| \; \tre' |a,b\rangle \otimes |c,d\rangle \\
= (-1)^{b-b'} \delta_{a,d'} \delta_{a',d} \delta_{b,-c} \delta_{b',-c'}.
\end{multline}
Its symmetrized version satisfies
\be
\label{tre}
\tre \; = -\tfrac12 \vec T_{x} \cdot \vec T_{y} - (\vec T_{x} \cdot \vec T_{y})^{2} + \tfrac32 \caS_{x,y}.
\ee
The correspondence between the transition operators and the parameters $J_{1},J_{2}$ is illustrated in Fig.\ \ref{fig phd loops}.
\bfig
\centerline{\includegraphics[width=55mm]{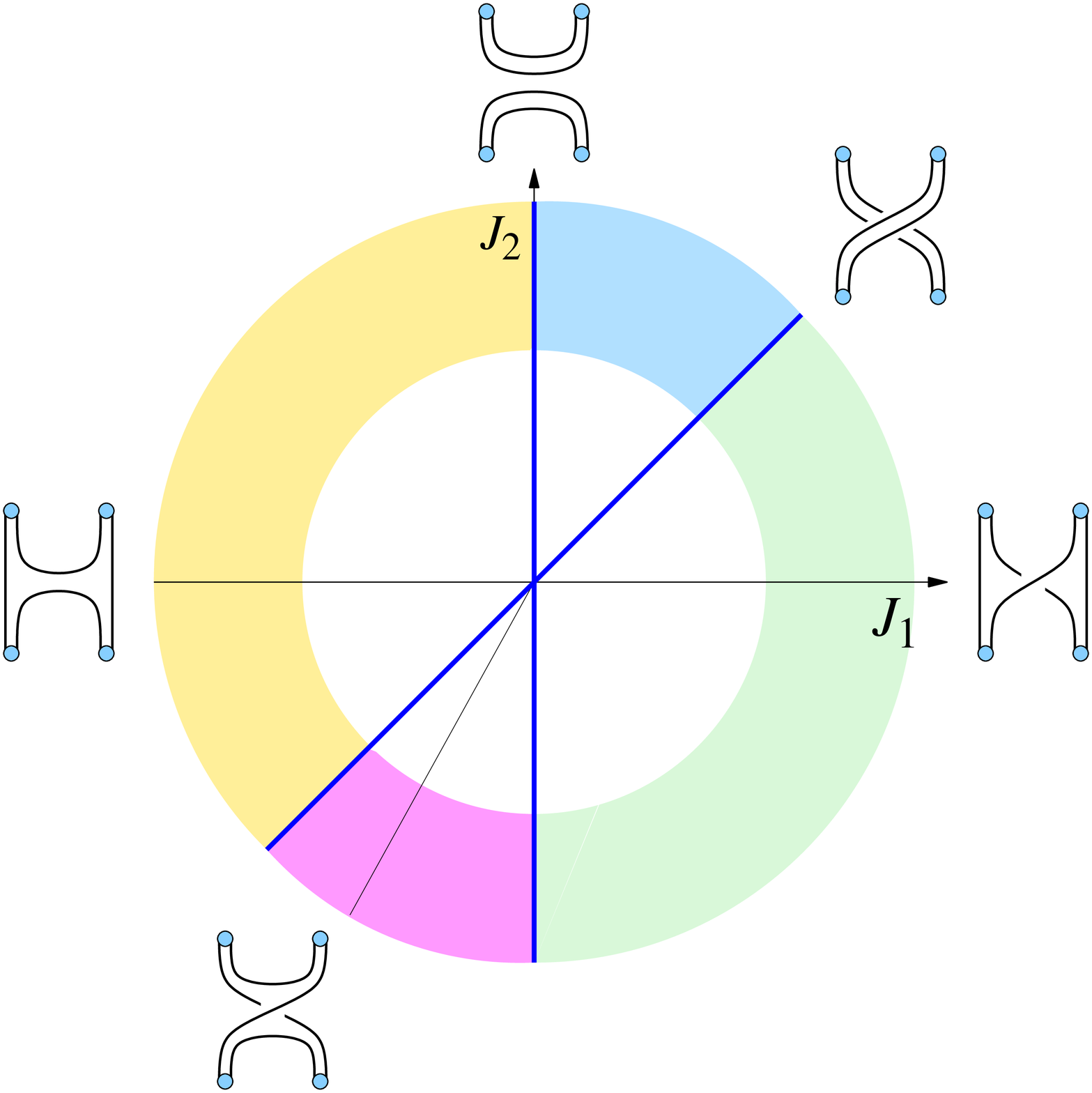}}
\caption{(Color online) The random loop representations for the different interactions. Convex combinations are possible. The loop representation in the direction $(J_{1}=-\frac12, J_{2}=-1)$, with transition operator \eqref{tre}, carries signs.}
\label{fig phd loops}
\efig
The Hamiltonian $H_{\Lambda}'$ can be written as linear combinations of the transition operators. In order to avoid future sign problems, we choose the following linear combinations. For $0 \leq J_{2} \leq J_{1}$,
\be
\label{Ham F}
H_{\Lambda}' = -\sum_{\langle x,y \rangle} \Bigl( 2(J_{1}-J_{2}) \;\tra + J_{2} \;\trb + (-J_{1}+2J_{2}) \caS_{x,y} \Bigr);
\ee
for $0 \leq J_{1} \leq J_{2}$,
\be
H_{\Lambda}' = -\sum_{\langle x,y \rangle} \Bigl( (J_{2}-J_{1}) \;\trc + J_{1} \;\trb + J_{2} \caS_{x,y} \Bigr);
\ee
and for $J_{1}<0, J_{2}\geq0$,
\be
\label{Ham AF}
H_{\Lambda}' = -\sum_{\langle x,y \rangle} \Bigl( -2J_{1} \;\trd + J_{2} \;\trc + (J_{1}+J_{2}) \caS_{x,y} \Bigr).
\ee

\subsection{Nachtergaele's loop representations}
\label{sec Nacht}

Let $|\sigma\rangle = \otimes_{x\in\Lambda} |\sigma_{x}^{(1)}, \sigma_{x}^{(2)} \rangle$, $\sigma_{x}^{(i)} = \pm\frac12$, denote the elements of the basis of $\caH_{\Lambda}'$ where all $(\sigma^{3} \otimes \bbone)_{x}$ and $(\bbone \otimes \sigma^{3})_{x}$ are diagonal. We can again use \eqref{Poisson exp} and we obtain an expansion that is similar to \eqref{full expansion}. Because of the properties of the transition operators, the expansion has graphical interpretation.

Let us examine the case of the Hamiltonian $H_{\Lambda}'$ for $0 \leq J_{2} \leq J_{1}$, given in Eq.\ \eqref{Ham F}. We can neglect the term $(-J_{1}+2S_{2}) \caS_{x,y}$ and add $2J_{1} + J_{2}$ times the identity so that we can use the Poisson expansion formula \eqref{Poisson exp}. Indeed, this may change the partition function, but it does not change the Gibbs states. A realization can then be represented as in Fig.\ \ref{fig st-conf}.
\bfig
\centerline{\includegraphics[width=70mm]{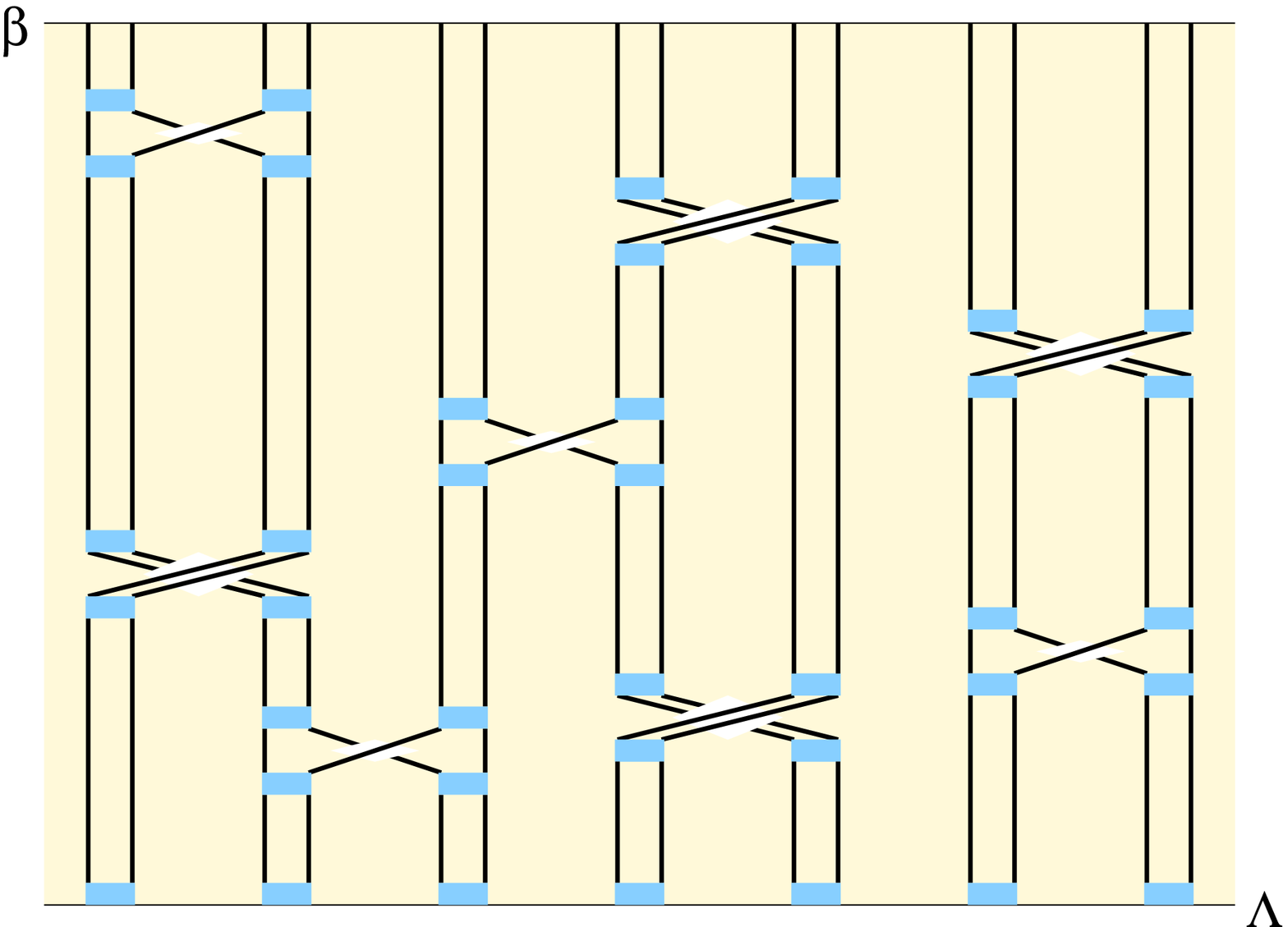}}
\caption{(Color online) A realization of the Poisson point process. The lattice is one-dimensional here but the representation applies to arbitrary dimensions.}
\label{fig st-conf}
\efig
There are two vertical lines on top of each site and neighboring lines are sometimes connected by the transitions. The little boxes represent the action of the symmetrization operator $\caS_{x}$ in Eq.\ \eqref{sym}, namely to average over an identity and a transposition. Graphically,

\smallskip
\centerline{\includegraphics[width=50mm]{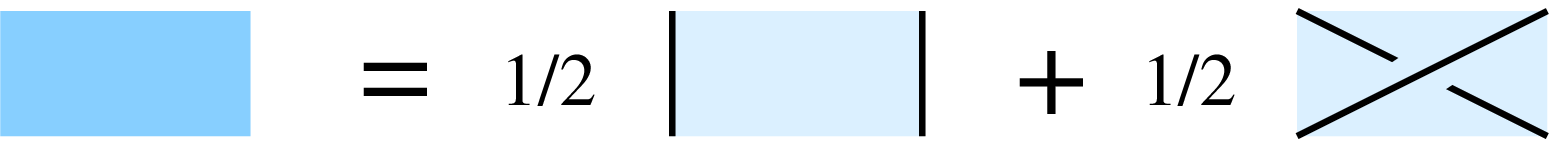}}

Let $\overline\omega$ denote the realization $\omega$ together with a choice of the status of all symmetrization boxes. We keep the notation $\rho$ for the corresponding measure.
Loop configurations are obtained by connecting the vertical lines according to the transition symbols. The sum over configurations $|\sigma_{1}\rangle, \dots, |\sigma_{k}\rangle$ amounts to a choice over assignments of a spin value $\pm\frac12$ to each loop. The partition function is then given by
\be
Z_{\Lambda}' = \int\rho(\dd\overline\omega) \; 2^{|\caL(\overline\omega)|},
\ee
where $|\caL(\overline\omega)|$ is the number of loops in the realization $\overline\omega$. A similar expansion can be performed for the correlation functions. We find
\be
\label{corr loops}
\Tr (\sigma^{3} \otimes \bbone)_{x} (\sigma^{3} \otimes \bbone)_{y} \e{-\beta H_{\Lambda}'} = \int\rho(\dd\overline\omega) \sum_{\sigma:\overline\omega} \sigma_{x}^{(1)} \sigma_{y}^{(1)}.
\ee
If the realization $\overline\omega$ does not connect the first lines of the sites $x$ and $y$, the sum gives zero. If the first lines of $x,y$ belong to the same loop, the sum gives $\frac14 2^{|\caL(\overline\omega)|}$. Recalling Eqs \eqref{T Pauli} and \eqref{corresp}, we get for $0 \leq J_{2} \leq J_{1}$:
\be
\label{spin-spin}
\langle S_{x}^{3} S_{y}^{3} \rangle_{H_{\Lambda}} = \langle T_{x}^{3} T_{y}^{3} \rangle_{H_{\Lambda}'} = \bbP \bigl( (x,1) \leftrightarrow (y,1) \bigr),
\ee
where the right side is the probability, with measure $\frac1{Z_{\Lambda}'} \dd\rho$, that the first lines of $x$ and $y$ belong to the same loop.

\bfig
\centerline{\includegraphics[width=70mm]{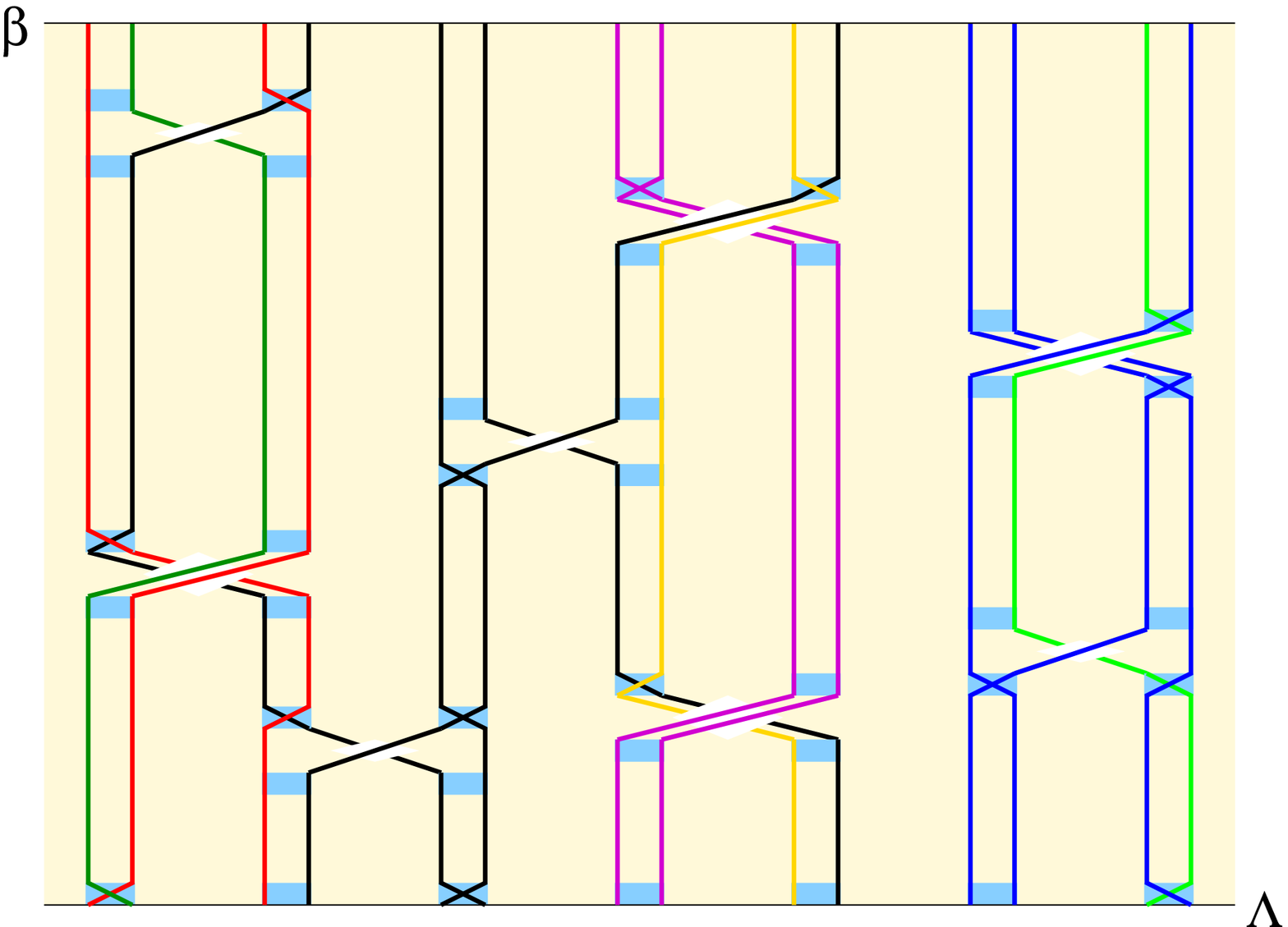}}
\caption{(Color online) The realization $\overline\omega$ and its loop configuration. Here, $|\caL(\overline\omega)|=7$.}
\label{fig st-loops}
\efig

The situation with transition operators $\trd$ and $\trc$ is similar but there are two important differences. Indeed, because of the matrix elements of Eqs \eqref{matrix elements}, we have that
\begin{itemize}
\item the spin value in the loop changes sign at those transitions where the vertical direction changes;
\item there are extra factors, namely $\e{\ii\pi a}$ for transitions of the form ${\lower 1mm \hbox{$\ulcorner\!\urcorner$}}$ and $\e{-\ii\pi a}$ for transitions of the form $\llcorner\!\lrcorner$, where $a=\pm\frac12$ is the value of the spin at sublattice A.
\end{itemize}
The factors give 1 when the lattice is bipartite and (i) the transitions $\trd$ and $\trc$ are between distinct sublattices only; (ii) the transpositions are between sites in the same sublattices. Thus $\trd$ and $\trc$ can be combined with one another, but they cannot be combined with $\tra$ and $\trb$. Actually, it is possible to combine $\trb$ and $\trc$ for the nematic phase. But in this case we can also use the simpler loop representation of Section \ref{sec simple loops} where each site has a single line.

The partition function for the Hamiltonian $H_{\Lambda}'$ in \eqref{Ham AF} is again $Z_{\Lambda}' = \int\rho(\dd\overline\omega) 2^{|\caL(\overline\omega)|}$. As for correlation functions, \eqref{corr loops} holds with the new notion of loops. The sum over $\sigma \! : \! \overline\omega$ gives (i) 0 if the first lines of $x,y$ do not belong to the same loop; (ii) $\frac14 2^{|\caL(\overline\omega)|}$ if they belong to the same loop and $x,y$ belong to the same sublattice; (iii) $-\frac14 2^{|\caL(\overline\omega)|}$ if they belong to the same loop and $x,y$ belong to different sublattices. Then we find for $J_{1}\leq0, J_{2}\geq0$:
\be
\langle S_{x}^{3} S_{y}^{3} \rangle_{H_{\Lambda}} = \langle T_{x}^{3} T_{y}^{3} \rangle_{H_{\Lambda}'} = (-1)^{x-y} \bbP \bigl( (x,1) \leftrightarrow (y,1) \bigr).
\ee

Let us conclude this subsection by giving two identities. A similar calculation as above gives $\langle (T_{x}^{3})^{2} \rangle_{H_{\Lambda}'} = \tfrac12 + \tfrac12 \bbP \bigl( (x,1) \leftrightarrow (x,2) \bigr)$. Since the first term is equal to $\frac23$, we find that
\be
\label{surprise}
\bbP \bigl( (x,1) \leftrightarrow (x,2) \bigr) = \tfrac13
\ee
for all values of $\beta$ and of interaction parameters. This should be understood as an effect of the symmetrization operators, which connect the two points with probability $1/3$ (if initially disconnected), or disconnect them with probability $2/3$ (if initially connected). The factor $2^{|\caL(\overline\omega)|}$ is important. Next, we compute the expression for the quadrupolar correlation function. Recall the operator $A_{x} = (S_{x}^{3})^{2} - \frac23$. We find
\be
\label{pas facile}
\begin{aligned}
\langle & A_{x} A_{y} \rangle_{H_{\Lambda}} = -\tfrac1{36} + \tfrac14 \bbP \biggl(
{\tiny\begin{tikzcd}
(x,1) \arrow[bend left]{d} & (y,1) \arrow[bend left]{d} \\
(x,2) \arrow[bend left]{u} & (y,2) \arrow[bend left]{u}
\end{tikzcd}}
\biggr) \\
&+ \tfrac12 \bbP \biggl(
{\tiny\begin{tikzcd}
(x,1) \arrow[bend left]{r} & (y,1) \arrow[bend left]{l} \\
(x,2) \arrow[bend left]{r} & (y,2) \arrow[bend left]{l}
\end{tikzcd}}
\biggr) \\
&+ \tfrac14 \bbP \Bigl(
(x,1), (x,2), (y,1), (y,2) \text{ in same loop}
\Bigr).
\end{aligned}
\ee
It seems quite equivalent to the spin-spin correlation function \eqref{spin-spin}, and it would be interesting to work out precise comparison. Notice that Eqs \eqref{surprise} and \eqref{pas facile} are valid in the whole region $J_{2}\geq0$.

\section{Joint distribution of the lengths of long loops}
\label{sec loop lengths}

\subsection{Macroscopic loops and random partitions}

Let $L^{(1)}, L^{(2)}, \dots$ denote the lengths of the loops in decreasing order. Here, the ``length'' of the loop is by definition the sum of lengths of its vertical components; thus $0 \leq L^{(i)} \leq \beta|\Lambda|$ and $\sum_{i} L^{(i)} = \beta |\Lambda|$.
We expect the largest ones to be of order of the volume, and the smaller ones to be of order of unity. We actually expect that the following limit exists for almost all realizations of loop configurations:
\be
\lim_{k\to\infty} \lim_{\Lambda \to \bbZ^{d}} \sum_{i=1}^{k} \frac{L^{(i)}}{\beta |\Lambda|} = m(\beta).
\ee
Further, we expect that the limit $m = m(\beta)$ takes a fixed value. The order of limits is of course important, the result is trivially 1 if the order is reversed. Next, in the limits $\Lambda \to \bbZ^{d}$ then $k\to\infty$, the sequence $\bigl( \frac{L^{(1)}}{\beta |\Lambda| m}, \frac{L^{(2)}}{\beta |\Lambda| m}, \dots \bigr)$ is a random partition of $[0,1]$. It turns out that its probability measure can be described explicitly; it is a Poisson-Dirichlet distribution PD($\theta$) for a suitable parameter $\theta>0$.

Aldous conjectured that PD(1) occurs in the random interchange model on the complete graph; this was proved by Schramm \cite{Sch}, who showed that the time evolution of the loop lengths is described by an effective split-merge process (or ``coagulation-fragmentation''). The random interchange model is equivalent to the random loop model of Section \ref{sec simple loops} without the factor $3^{|\caL(\omega)|}$. It was later understood that the Poisson-Dirichlet distribution is also present in three-dimensional systems \cite{GUW}. This behavior is actually fairly general and concerns many systems where one-dimensional objects have macroscopic size. The distribution PD(1) has been confirmed numerically in a model of lattice permutations \cite{GLU} and analytically in the related annealed model \cite{BU}. More general PD($\theta$) have been confirmed numerically in O($n$) loop models \cite{NCSOS}.

There are several definitions of Poisson-Dirichlet distributions. The simplest one is through the Griffiths-Engen-McCloskey (GEM) distribution. Let $X_{1}, X_{2}, \dots$ be independent identically distributed Beta($\theta$) random variables. That is, each $X_{i}$ takes value in $[0,1]$ and $\bbP(X>s) = (1-s)^{\theta}$ for $0 \leq s \leq 1$. The following is a random partition of $[0,1]$:
\[
\Bigl( X_{1}, (1-X_{1}) X_{2}, (1-X_{1}) (1-X_{2}) X_{3}, \dots \Bigr).
\]
The corresponding distribution is called GEM$(\theta)$. After reordering the numbers in decreasing order, the resulting random partition has PD($\theta$) distribution.

The probability that two random numbers, chosen uniformly in $[0,1]$, belong to the same element in PD($\theta$), can be calculated with GEM($\theta$). They both belong to the $k$th element with probability
\be
\bbE \bigl( (1-X_{1})^{2} \dots (1-X_{k-1})^{2} X_{k}^{2} \bigr) = \bbE \bigl( (1-X_{1})^{2} \bigr)^{k-1} \bbE \bigl( X_{1}^{2} \bigr).
\ee
We have $\bbE \bigl( (1-X_{1})^{2} \bigr) = \theta / (\theta+2)$ and $\bbE \bigl( X_{1}^{2} \bigr) = 2 / (\theta+1) (\theta+2)$. Summing over $k$, we find that the probability that two random numbers belong to the same element is equal to $1/(\theta+1)$.

\subsection{Effective split-merge process}

The strategy is to view the random loop measure as the stationary measure of a suitable stochastic process. The effective process on random partitions is a split-merge process with suitable rates. Then its invariant measure is a PD$(\theta$) distribution with $\theta$ that depends on the rates.

We consider the simpler loop model of Section \ref{sec simple loops} and explain later that modifications are minimal for the other loop models. Let $R(\omega,\omega')$ be the transition matrix of the stochastic process, that gives the rate at which $\omega'$ occurs when the configuration is $\omega$. With $C(\omega)$ the number of crosses and $B(\omega)$ the number of double bars, the detailed balance property is
\bm
3^{|\caL(\omega)|} (u \dd t)^{C(\omega)} \bigl( (1-u) \dd t \bigr)^{B(\omega)} R(\omega,\omega') \\
= 3^{|\caL(\omega')|} (u \dd t)^{C(\omega')} \bigl( (1-u) \dd t \bigr)^{B(\omega')} R(\omega',\omega).
\end{multline}
We have discretized the interval $[0,\beta]$ with mesh $\dd t$. The following process satisfies the detailed balance property.
\begin{itemize}
\item A new cross appears in the interval $\{x,y\} \times [t,\dd t]$ at rate $\sqrt3 \, u \, \dd t$ if its appearance causes a loop to split; at rate $(u / \sqrt3) \, \dd t$ if its appearance causes two loops to merge; and at rate $u \, \dd t$ if its appearance does not modify the number of loops.
\item Same with bars, but with rate $1-u$ instead of $u$.
\item An existing cross and double bar is removed at rate $\sqrt3$ if its removal causes a loop to split; at rate $1/\sqrt3$ if its removal causes two loops to merge; and at rate 1 if the number of loops remains constant.
\end{itemize}
Notice that any new cross or double bar between two loops causes them to merge. When $u=1$, any new cross within a loop causes it to split.

Let $\gamma, \gamma'$ be two macroscopic loops of respective lengths $L,L'$. They are spread all over $\Lambda$ and they interact between one another, and among themselves, in an essentially mean-field fashion. There exists a constant $c_{1}$ such that a new cross or double bar that causes $\gamma$ to split, appears at rate $\tfrac12 \sqrt3 \, c_{1} \frac{L^{2}}{\beta |\Lambda|}$; a new cross or double bar that causes $\gamma$ and $\gamma'$ to merge appears at rate $(c_{1} / \sqrt3) \frac{L L'}{\beta |\Lambda|}$. There exists another constant $c_{2}$ such that the rate for an existing cross or double bar to disappear is $\tfrac12 \sqrt3 \, c_{2}  \frac{L^{2}}{\beta |\Lambda|}$ if $\gamma$ is split, and $(c_{2} / \sqrt3) \frac{L L'}{\beta |\Lambda|}$ if $\gamma$ and $\gamma'$ are merged. Consequently, $\gamma$ splits at rate
\be
\tfrac12 \sqrt3 (c_{1}+c_{2}) \frac{L^{2}}{\beta |\Lambda|} \equiv \tfrac12 r_{\rm s} L^{2}
\ee
and $\gamma, \gamma'$ merge at rate
\be
\frac1{\sqrt3} (c_{1}+c_{2}) \frac{L L'}{\beta |\Lambda|} \equiv r_{\rm m} L L'.
\ee
Because of effective averaging over the whole domain, the constants $c_{1}$ and $c_{2}$ are the same for all loops and for both the split and merge events. This key property is certainly not obvious and the interested reader is referred to a detailed discussion for lattice permutations with numerical checks \cite{GLU}. It follows that the lengths of macroscopic loops satisfy an effective split-merge process, and the invariant distribution is Poisson-Dirichlet with parameter $\theta = r_{\rm s} / r_{\rm m} = 3$ \cite{Bertoin,GUW}.

The case $u \in (0,1)$ is different because loops split with only half the rate above. Indeed, the appearance of a new transition within the loop may just rearrange it: topologically, this is like $0 \leftrightarrow 8$, see Fig.\ \ref{fig 0-8} for illustration. This yields PD$(\frac32)$. Notice that this cannot happen when $u=1$, or when $u=0$ on a bipartite graph.
\bfig
\centerline{\includegraphics[width=70mm]{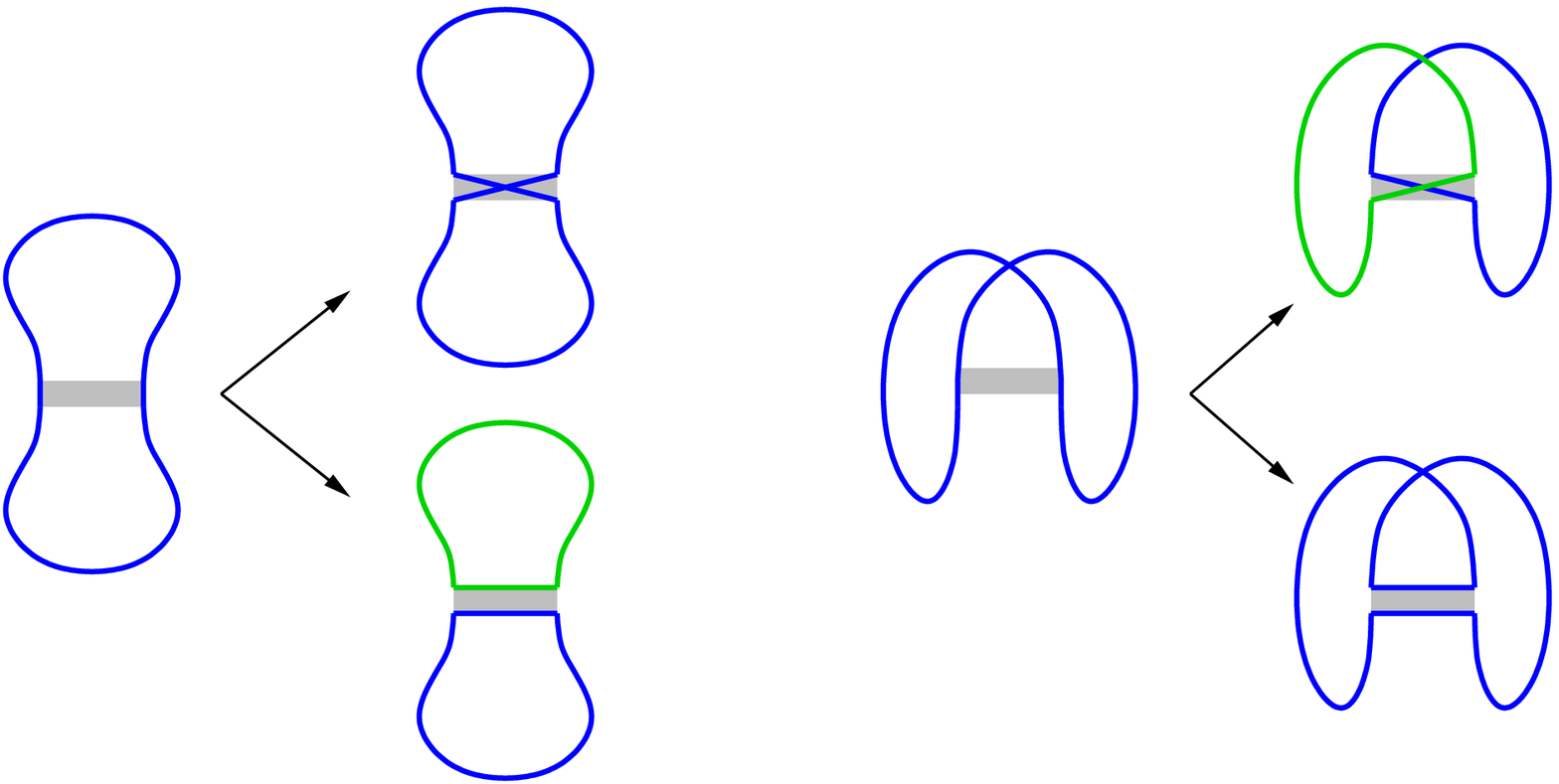}}
\caption{(Color online) When $u \in (0,1)$, a local change involving two legs of the same loop may rearrange it rather than split it. This figure shows all cases corresponding to the addition of a transition. The loop necessarily splits when $u=1$; and also when $u=0$ on a bipartite graph.}
\label{fig 0-8}
\efig

It follows from these considerations that the macroscopic loops of the model described in Section \ref{sec simple loops} have the following Poisson-Dirichlet distributions:
\begin{itemize}
\item PD(3) when $J_{1} = J_{2} >0$ and when $J_{1}=0, J_{2}>0$;
\item PD($\frac32$) when $0 < J_{1} < J_{2}$.
\end{itemize}

If the graph is not bipartite, it is still PD(3) when $J_{1} = J_{2} >0$; but it is PD($\frac32$) otherwise, including when $J_{1}=0, J_{2}>0$.

The heuristics for the more complicated loop models of Subsection \ref{sec Nacht} is similar. It helps to reformulate the setting, by considering transitions that connect any vertical lines; this allows to neglect the symmetrization boxes (except at time 0). The discussion for transitions $\tra$ or $\trd$ is then as before and the distribution is PD(2).

The transitions $\trb$ and $\trc$ have multiple connections and the argument is a bit more difficult. But it is possible to obtain an effective split-merge process where two changes occur in succession, with the appropriate rates. Hence PD(2) again. It is essential for the heuristics that, if two vertical lines at the same site belong to macroscopic loops, they are uncorrelated. This seems to fail because of \eqref{surprise}. But this holds since symmetrization boxes are discarded.

This heuristics {\it does not} apply if the model has only transitions $\trb$ (or $\trc$). Indeed, there are non-trivial correlations between lines at the same site. It should be possible to modify the argument so as to identify the correct Poisson-Dirichlet. But this is not necessary since we can use the simpler loop model of Section \ref{sec simple loops}.

\subsection{Long-range correlation in models of random loops}
\label{sec justifications}

At low temperatures, our large system contains many loops, that are either macroscopic or microscopic. (Mesoscopic loops have vanishing density.) While the lengths of macroscopic loops varies, their total density is essentially fixed and is denoted $m(\beta)$. The probability that two distant points belong to the same loop is then equal to the probability that they both belong to macroscopic loops $m(\beta)^{2}$, times the probability that they belong to the same partition element. If the random partition has Poisson-Dirichlet distribution with parameter $\theta$, this probability is $1/(\theta+1)$ and we find therefore
\be
\label{long corr}
\bbP( x \leftrightarrow y ) = \frac{m(\beta)^{2}}{\theta+1}.
\ee
Let us insist that this equation is meant as an identity in the limits $\Lambda\to\bbZ^{d}$ and $\|x-y\|\to\infty$.

The recent article of V\"oll and Wessel on the ground state of the model in the triangular lattice \cite{VW} allows us to confront the above heuristics with numerical results. Namely, the black curve in their Fig.\ 1 is proportional to the length of the loop that contains a given point. At $J_{1}=J_{2}$ (this corresponds to $\vartheta = -\frac34 \pi$ in \cite{VW}) it is equal to $m(\beta) / (3+1)$; for $J_{1} \leq J_{2}$ (larger $\vartheta$ in \cite{VW}) it is equal to $m(\beta) / (\frac32 + 1)$. The discontinuity of the black curve at $J_{1}=J_{2}$ is characterized by the ratio of the above values, i.e.\ 8/5. The numerical data from the figure gives a ratio that is roughly equal to 2.6/1.6, which is indeed remarkably close.

We now calculate the long-distance correlations in the various phases.
In the ferromagnetic phase, we have $\theta=2$. Combining Eqs \eqref{spin-spin} and \eqref{long corr}, we get that the left side of \eqref{calcul ferro} is equal to
\be
\langle S_{x}^{3} S_{y}^{3} \rangle_{H_{\Lambda}} = \bbP \bigl( (x,1) \leftrightarrow (y,1) \bigr) \; \substack{\Lambda\to\bbZ^{d} \\ \xrightarrow{\hspace*{13mm}} \\ \|x-y\|\to\infty} \; \tfrac13 m(\beta)^{2}.
\ee
We need an expression for the magnetization of extremal states. It is possible to add an external field of the form $\sum_{y} T_{y}^{3}$ to the Hamiltonian $H_{\Lambda}'$ and to perform a similar expansion, using Trotter or Duhamel formul\ae. With $h$ positive but tiny, we get that small loops can still carry any spin values, but long loops must have spin value $+\frac12$. Consequently,
\be
\begin{split}
\langle S_{x}^{3} \rangle_{\vec e_{3}}^{\rm ferro} &= \lim_{h\to0+} \lim_{\Lambda\to\bbZ^{d}} \langle T_{x}^{3} \rangle_{H_{\Lambda}' - h \sum_{y} T_{y}^{3}} \\
&= \lim_{h\to0+} \lim_{\Lambda\to\bbZ^{d}} \langle (2\sigma^{3} \otimes \bbone)_{x} \rangle_{H_{\Lambda}' - h \sum_{y} T_{y}^{3}} \\
&= m(\beta).
\end{split}
\ee
The last line in Eq.\ \eqref{calcul ferro} is then also equal to $\frac13 m(\beta)^{2}$. This confirms the decomposition \eqref{ferro decomposition}.

Next, we verify the SU(3) decomposition \eqref{decomposition SU(3)}. For $J_{1}=J_{2}$, we use the simpler loop representation of Section \ref{sec simple loops}. Let $m(\beta)$ be the fraction of sites in macroscopic loops. The distribution is PD(3), so the left side of Eq.\ \eqref{calcul SU(3)} is
\be
\langle S_{x}^{3} S_{y}^{3} \rangle_{H_{\Lambda}} = \tfrac23 \bbP(x \leftrightarrow y) \; \substack{\Lambda\to\bbZ^{d} \\ \xrightarrow{\hspace*{13mm}} \\ \|x-y\|\to\infty} \;  \tfrac16 m(\beta)^{2}.
\ee
Notice that the same calculation gives
\be
\langle \lambda_{x}^{3} \lambda_{y}^{3} \rangle_{H_{\Lambda}} = \langle \lambda_{x}^{8} \lambda_{y}^{8} \rangle_{H_{\Lambda}} = \tfrac23 \bbP(x \leftrightarrow y),
\ee
which justifies the first identity in \eqref{calcul SU(3)}.
In the extremal state $\langle \cdot \rangle_{\vec e_{3}}$, long loops have spin value 1 and small loops  have any value $-1,0,+1$. It follows that
\be
\langle \lambda_{x}^{3} \rangle_{\vec e_{3}} = m(\beta), \qquad \langle \lambda_{x}^{8} \rangle_{\vec e_{3}} = \tfrac1{\sqrt3} m(\beta).
\ee
The last line of \eqref{calcul SU(3)} is equal to $\frac16 m(\beta)^{2}$; this confirms the SU(3) decomposition \eqref{decomposition SU(3)}.

Next is the spin nematic phase, for $0 \leq J_{1} \leq J_{2}$. We again use the simpler loop representation; the correct distribution is now PD($\frac32$). The quadrupolar correlation function in the symmetric Gibbs state gives
\be
\langle A_{x} A_{y} \rangle_{H_{\Lambda}} = \tfrac29 \bbP(x \leftrightarrow y) \; \substack{\Lambda\to\bbZ^{d} \\ \xrightarrow{\hspace*{13mm}} \\ \|x-y\|\to\infty} \;  \tfrac4{45} m(\beta)^{2}.
\ee
This the left side of Eq.\ \eqref{calcul nem}.
The spin values of the long loops of extremal Gibbs states is 0. Thus
\be
\langle (S_{x}^{i})^{2} - \tfrac23 \rangle_{\vec e_{3}}^{\rm nem} = \begin{cases} \phantom{-} \frac13 m(\beta) & \text{if } i=1,2, \\ -\frac23 m(\beta) & \text{if } i=3. \end{cases}
\ee
This implies that the right side of \eqref{calcul nem} is also equal to $\frac4{45} m(\beta)^{2}$. This confirms that the extremal nematic states are given by Eq.\ \eqref{extremal nem state}, and that the symmetric Gibbs state has the decomposition \eqref{decomposition nem}.

The justification of the staggered SU(3) decomposition, Eq.\ \eqref{decomposition stag SU(3)}, is identical to that of the SU(3) decomposition. And the justification for the antiferromagnetic decomposition \eqref{decomposition AF} is also identical to that of the ferromagnetic decomposition.

To summarize, we have given a precise characterization of the symmetry breakings that occur in a spin 1 quantum model at low temperatures and in dimensions three and higher. We have used random loop representations to confirm the decompositions into extremal Gibbs states. While the present discussion is not rigorous, it is expected to be exact. We have only considered the case $J_{2}\geq0$; a difficult challenge is to clarify the phase diagram for $J_{2}<0$, and in particular the staggered spin nematic phase.

\medskip {\small {\bf Acknowledgments:} It is a pleasure to thank J\"urg Fr\"ohlich and Gian Michele Graf for long discussions and for key suggestions. I am also grateful to Marek Biskup, John Chalker, Sacha Friedli, Roman Koteck\'y, Benjamin Lees, Nicolas Macris, Charles-\'Edouard Pfister, Tom Spencer, and Yvan Velenik, for discussions and encouragements. The exposition benefitted from useful comments of the two referees.
}

\end{document}